\documentclass[12pt,english]{paper}
\usepackage[T1]{fontenc}
\usepackage[latin1]{inputenc}
\usepackage{graphicx}

\makeatletter

\makeatletter

\makeatletter

\makeatletter

\usepackage{geometry}

\makeatletter

\usepackage{epsfig}

\makeatother

\makeatother

\makeatother

\makeatother

\usepackage{babel}
\makeatother
\begin{document}

\title{Application of Bayesian Neural Networks to Energy Reconstruction in EAS Experiments for ground-based TeV Astrophysics }

\author{Ying Bai$^1$, Ye Xu$^{1,2}$%
\thanks{Corresponding author, e-mail address: xuy@fjut.edu.cn%
}, Jia Pan$^1$, JieQin Lan$^1$, WeiWei Gao$^1$}

\maketitle
$^1$School of Mathematics and Physics, Fujian University of Technology, Fuzhou 350118, China
\par
$^2$School of Information Science and Engineering, Fujian University of Technology, Fuzhou 350118, China

\begin{abstract}
A toy detector array is designed to detect a shower generated by the interaction between a TeV cosmic ray
and the atmosphere. In the present paper, the primary energies of showers detected by the detector array are
reconstructed with the algorithm of Bayesian neural networks (BNNs) and
a standard method like the LHAASO experiment \cite{lhaaso-ma},
respectively. Compared to the standard method, the energy resolutions are significantly improved using the BNNs. And the
improvement is more obvious for the high energy showers than the
low energy ones.
\end{abstract}
\begin{flushleft}
\par\end{flushleft}

\section{Introduction}
TeV astrophysics has become the window of extragalactic
science by about three decades of development. Topics to which
ground-based observations of TeV cosmic rays can make very important contributions
are the following: supermassive black hole, acceleration mechanism of
cosmic ray, dark matter and so on.
\par
Extended Air Shower (EAS from now on) arrays \cite{hawc,asgamma,lhaaso} and Imaging Atmospheric Cherenkov
Telescopes (IACTs from now on) \cite{hess,magic,veritas} are two important techniques for ground-based observations of TeV
cosmic rays. Compared to an IACT, a large EAS array can provide the large
field of view(about 90 deg; 1.8 sr) and nearly 100$\%$ duty cycle. So these
characteristics make their observatories particularly suited to conduct
all-sky surveys and detect emission from extended astrophysical sources
(larger than about 1 deg, e.g. plane of the Galaxy) \cite{techwhite}.
\par
The primary energy of a shower can be evaluated by the lateral density of observed shower particles in an EAS observatory.
And the particle density at a certain distance from the cascade core is used as an energy estimator \cite{lhaaso-ma}.
Whereas the energy resolution evaluated by this method can reach only about 18$\%$
at 1 PeV \cite{lhaaso-ma}.
\par
The Bayesian neural networks (BNNs from now on) \cite{rmneal} is an algorithm
of the neural networks trained by the Bayesian statistics. It is not only a non-linear
function, but also controls model complexity. So its flexibility makes it possible to
discover more general relationships in data than the conventional statistical methods
and its preferring simple models make it possible to solve the over-fitting problem
better than the general neural networks \cite{rbeale}. Compared to the
energy estimator mentioned above, the BNNs is more suitable for the energy reconstruction in EAS array experiments. In the
present paper, an EAS due to the interaction between a proton and the atmosphere is simulated using the Aires2.8.2a with SIBYLL package \cite{aires}.
The BNNs is applied to reconstruct the primary energy of a shower. It is discussed the comparison between the results using the BNNs and
energy estimator in the following sections.

\section{The Regression with BNNs \cite{rmneal,pcbhat}}

The idea of the BNNs is to regard the process of training a neural network
as a Bayesian inference. Bayes' theorem is used to assign a posterior
density to each point, $\bar{\theta}$, in the parameter space of
the neural networks. Each point $\bar{\theta}$ denotes a neural network.
In the method of BNNs, one performs a weighted average over all points
in the parameter space of the neural network, that is, all neural
networks. The methods make use of training data \{($x_{1}$,$t_{1}$),
($x_{2}$,$t_{2}$),...,($x_{n}$,$t_{n}$)\}, where $t_{i}$ is the
known target value associated with data $x_{i}$, which has $P$ components
if there are $P$ input values in the regression. That is the set
of data $x=$($x_{1}$,$x_{2}$,...,$x_{n}$) which corresponds to
the set of target $t=$($t_{1}$,$t_{2}$,...,$t_{n}$). The posterior
density assigned to the point $\bar{\theta}$, that is, to a neural
network, is given by Bayes' theorem

\begin{center}
\begin{equation}
p\left(\bar{\theta}\mid x,t\right)=\frac{\mathit{p\left(x,t\mid\bar{\theta}\right)p\left(\bar{\theta}\right)}}{p\left(x,t\right)}=\frac{p\left(t\mid x,\bar{\theta}\right)p\left(x\mid\bar{\theta}\right)p\left(\bar{\theta}\right)}{p\left(t\mid x\right)p\left(x\right)}=\frac{\mathit{p\left(t\mid x,\bar{\theta}\right)p\left(\bar{\theta}\right)}}{p\left(t\mid x\right)}\end{equation}

\par\end{center}

\begin{flushleft}
where data $x$ does not depend on $\bar{\theta}$, so $p\left(x\mid\theta\right)=p\left(x\right)$.
We need the likelihood $p\left(t\mid x,\bar{\theta}\right)$ and the
prior density $p\left(\bar{\theta}\right)$, in order to assign the
posterior density $p\left(\bar{\theta}\mid x,t\right)$to a neural
network defined by the point $\bar{\theta}$. $p\left(t\mid x\right)$
is called evidence and plays the role of a normalizing constant, so
we ignore the evidence.
\end{flushleft}
\begin{flushleft}
We consider a class of neural networks defined by the function
\par\end{flushleft}

\begin{center}
\begin{equation}
y\left(x,\bar{\theta}\right)=b+{\textstyle {\displaystyle \sum_{j=1}^{H}v_{j}\tanh\left(a_{j}+\sum_{i=1}^{P}u_{ij}x_{i}\right)}}\end{equation}

\par\end{center}

\begin{flushleft}
The neural networks have $P$ inputs, a single hidden layer of $H$
hidden nodes and one output. Because of quicker convergence and better accuracy,
 a tanh function is used as an activation function for the hidden layer of the BNNs in our work (See Eq.(2)). In the particular BNNs described here,
each neural network has the same structure. The parameter $u_{ij}$
and $v_{j}$ are called the weights and $a_{j}$ and $b$ are called
the biases. Both sets of parameters are generally referred to collectively
as the weights of the BNNs, $\bar{\theta}$. $y\left(x,\bar{\theta}\right)$
is the predicted target value. We assume that the noise on target
values can be modeled by the Gaussian distribution. So the likelihood
of $n$ training events is
\par\end{flushleft}

\begin{center}
\begin{equation}
p\left(t\mid x,\bar{\theta}\right)=\prod_{i=1}^{n}\exp[-(t_{i}-y\left(x_{i},\bar{\theta}\right))^{2}/2\sigma^{2}]=\exp[-\sum_{i=1}^{n}(t_{i}-y\left(x_{i},\bar{\theta}\right))^2/2\sigma^{2})]\end{equation}

\par\end{center}

\begin{flushleft}
where $t_{i}$ is the target value, and $\sigma$ is the standard
deviation of the noise. It has been assumed that the events are independent
with each other. Then, the likelihood of the predicted target value
is computed by Eq. (3).
\par\end{flushleft}

\begin{flushleft}
We get the likelihood, meanwhile we need the prior to compute the
posterior density. But the choice of prior is not obvious.
Experience suggests a reasonable class is the priors of Gaussian class
centered at zero, which prefers smaller rather than larger weights,
because smaller weights yield smoother fits to data . In the paper,
a Gaussian prior is specified for each weight using the Bayesian neural
networks package of Radford Neal%
\footnote{R. M. Neal, \emph{Software for Flexible Bayesian Modeling and Markov
Chain Sampling}, http://www.cs.utoronto.ca/\textasciitilde{}radford/fbm.software.html%
}. The variance for weights belonging to a given group(either
input-to-hidden weights($u_{ij}$), hidden -biases($a_{j}$), hidden-to-output
weights($v_{j}$) or output-biases($b$)) is chosen to be the same:
$\sigma_{u}^{2}$, $\sigma_{a}^{2}$, $\sigma_{v}^{2}$, $\sigma_{b}^{2}$,
respectively. Since we don't know, a priori, what these variances
should be, their values are allowed to vary over a large range, while
favoring small variances. This is done by assigning each variance
a gamma prior
\par\end{flushleft}

\begin{center}
\begin{equation}
p\left(z\right)=\left(\frac{\alpha}{\mu}\right)^{\alpha}\frac{z^{\alpha-1}e^{-z\frac{\alpha}{\mu}}}{\Gamma\left(\alpha\right)}\end{equation}

\par\end{center}

\begin{flushleft}
where $z=\sigma^{-2}$, and with the mean $\mu$ and shape parameter
$\alpha$ set to some fixed plausible values. The gamma prior is referred
to as a hyperprior and the parameter of the hyperprior is called a
hyperparameter.
\par\end{flushleft}

\begin{flushleft}
Then, the posterior density, $p\left(\bar{\theta}\mid x,t\right)$,
is gotten according to Eqs. (3) and the prior of Gaussian distribution.
Given an event with data $x'$, an estimate of the target value is
given by the weighted average
\par\end{flushleft}

\begin{center}
\begin{equation}
\bar{y}\left(x'|x,t\right)=\int y\left(x',\bar{\theta}\right)p\left(\bar{\theta}\mid x,t\right)d\bar{\theta}\end{equation}

\par\end{center}

\begin{flushleft}
Currently, the only way to perform the high dimensional integral in
Eq. (5) is to sample the density $p\left(\bar{\theta}\mid x,t\right)$
with the Markov Chain Monte Carlo (MCMC) method\cite{rmneal,sabd,ma,pb}.
In the MCMC method, one steps through the $\bar{\theta}$ parameter
space in such a way that points are visited with a probability proportional
to the posterior density, $p\left(\bar{\theta}\mid x,t\right)$. Points
where $p\left(\bar{\theta}\mid x,t\right)$ is large will be visited
more often than points where $p\left(\bar{\theta}\mid x,t\right)$
is small.
\par\end{flushleft}

\begin{flushleft}
Eq. (5) approximates the integral using the average
\par\end{flushleft}

\begin{center}
\begin{equation}
\bar{y}\left(x'\mid x,t\right)\approx\frac{1}{L}\sum_{i=1}^{L}y\left(x',\bar{\theta_{i}}\right)\end{equation}

\par\end{center}

\begin{flushleft}
where $L$ is the number of points $\bar{\theta}$ sampled from $p\left(\bar{\theta}\mid x,t\right)$.
Each point $\bar{\theta}$ corresponds to a different neural network
with the same structure. So the average is an average over neural
networks, and is closer to the real value of $\bar{y}\left(x'\mid x,t\right)$,
when $L$ is sufficiently large.
\par\end{flushleft}

\section{Toy Detector Array and Simulation}
In our work, a toy detector array is designed to
detect TeV cosmic rays. And it is located on the ground with the altitude of 4300m, like Yangbajing, Tibet, China.
There are two components in the array:
one is an electron detector (ED from now on) array covering one km$^2$ square; The
other is a muon detector (MD from now on) array included in the ED array.
These detectors (EDs and MDs) consist of scintillator coun.
But each ED is a 1$\times$1 m$^2$ scintillator counter and covered
by one 0.5cm thick Pb plate used as a gamma converter. 3969 of them are located
in a square grid with a side length of 16 meters. Each MD is a 6$\times$6 m$^2$
scintillator counter and covered by overburden
of 3 meters thick earth to remove electro-magnetic components in
an air shower with a threshold energy of 2 GeV. 676 of them are located in a square grid with a side length of 39 meters.
\par
A shower due to the interaction between a proton and the atmosphere is simulated
by the Aires2.8.4a with SIBYLL package and the response of its secondary particles
deposited in the toy detector array is simulated with the GEANT4 package \cite{geant4}. The primary energy of an EAS is sampled
from a spectrum between 10 TeV to 10$^4$ TeV and its arrival directions is sampled from a isotropic distribution in a range of the zenith angle from 0 to $\pi$/4 and azimuth
angle from 0 to 2$\pi$.

\section{Primary Energy Reconstruction}
The lateral distribution of charged particles from an EAS has a direct relationship with its primary energy. The lateral distributions of electrons and muons are simulated with the Aires2.8.4a with SIBYLL package. To estimate the primary energy of an EAS, one has to observe lateral densities of the charged particles with ground detectors. In our work, $N_e$, $N_\mu$ and $N_{e\mu}$ are regarded as the lateral densities of the charged particles. $N_e$ and $N_\mu$ are the sums of electrons and muons in EDs and MDs in $R_c=50-100 m$ divided by $\cos{\theta}$, respectively. $R_c$ is the distance from hit to the shower core and $\theta$ is the zenith angle of the shower. $N_{e\mu}$ is the same parameter in Ref. \cite{lhaaso-ma}:
\par
\begin{center}
\begin{equation}
N_{e\mu} = \sqrt{N_{e}N_\mu}
\end{equation}
\end{center}
\par
In the present paper, the primary energy of an EAS is estimated with the linear fitting method(LFM from now on), like the LHAASO experiment \cite{lhaaso-ma}, and the BNNs, respectively. The parameters mentioned above are used as the inputs of the LFM and BNNs in our energy estimation.

\subsection{Primary Energy Reconstruction with LFM}
There is an approximative linear relationship between the logarithms of the primary energy and density of charged particles for an EAS. This relationship is used to estimate the primary energy of a shower in the LHAASO experiment \cite{lhaaso-ma}. In our work, the LFM by taken as a standard method is used to estimate the primary energy, too.
\par
The primary energy $E_0$ of a shower is sampled from a power law spectrum between 10 TeV to $10^4$ TeV whose index is set to -2.7 in our work. The $10^5$ events are generated by the Aires2.8.4a with SIBYLL package and used as the training sample for the LFM. That is, the approximative linear relationship between $\log{E_0}$ and $\log{N_{e\mu}}$ is obtained through fitting the training sample. Fig. 1 shows that this relationship:
\par
\begin{center}
\begin{equation}
\log{E_0}=a\log{N_{e\mu}}+b
\end{equation}
\end{center}
\par
The values of a and b are obtained after fitting the equation (8) to the training sample, respectively(see Fig. 1). They are 0.977 and 12.763, respectively.
We get the following equation:
\par
\begin{center}
\begin{equation}
E_0\;=\;5.794N_{e\mu}^{0.977}\ (\rm{TeV})
\end{equation}
\end{center}
\par
3000 showers are generated at 100 TeV with the Aires2.8.4a with SIBYLL package. The 8 different data files are prepared at 1000 TeV interval in the energy range from 1000 TeV to 8000 TeV in the same way. Besides, the 9 different data files are prepared at 5$^0$ interval in the zenith angle range from 3$^0$ to 45$^0$ at 3000 TeV in the same way.  These data files are taken as the test samples for the LFM, and their energies are reconstructed with the equation (9). They are used as the test samples for the BNN, too.
\par
In Fig. 2, the black triangles denote the energy resolutions reconstructed with the LFM at nine different energies.
In Fig. 3, the black triangles denote the energy resolutions reconstructed with the LFM at nine different zenith angles at 3000 TeV.

\subsection{Primary Energy Reconstruction with BNNs}
The primary energy of an EAS is sampled
from a spectrum between 10 TeV to 10$^4$ TeV. This spectrum is made up of three ranges with
uniform distributions. 5000, 5$\times$$10^4$ and 5$\times$$10^5$ showers are produced in the three ranges between
10 TeV and 100 TeV, 100 TeV and 1000 TeV, and 1000 TeV and 10$^4$ TeV, respectively. These showers are taken as the training
samples for the BNNs. The test samples for the BNNs are the same ones as the LFM.
In our work, $\log{(N_e+N_\mu)}$ and $\log{N_{e\mu}}$
are used as inputs to the BNNs, which have the input layer of 2 inputs,
the single hidden layer of 4 nodes and the output layer of a output. The logarithms of primary energies ($\log{E_0}$) for
the test samples are predicted by the BNNs. A Markov
chain of neural networks is generated using the Bayesian neural networks
package of Radford M. Neal \cite{rmn}, with the training sample, in the process
of the energy reconstruction. Seven hundred iterations, of twenty Markov chain
Monte-Carlo steps each, are used in our work. The neural network parameters are
stored after each iteration, since the correlation between adjacent
steps is very high. That is, the points in neural network parameter
space are saved to lessen the correlation after twenty steps. It is
also necessary to discard the initial part of the Markov chain because
the correlation between the initial point of the chain and the ones
of the part is very high. The initial three hundred iterations are
discarded in our work.
\par
In Fig. 2, the black squares denote the energy resolutions reconstructed with the BNNs at nine different energies.
In Fig. 3, the black squares denote the energy resolutions reconstructed with the BNNs at nine different zenith angles at 3000 TeV.

\section{Conclusion}

\begin{flushleft}
Fig. 2 and Fig. 3 illustrate the results of the energy
reconstruction with the LFM and BNNs. Fig. 2 shows that the resolutions of the
different primary energies reconstructed with the BNNs are
obviously different from the LFM. Compared to the LFM, the resolutions with the BNNs
decrease by 28.2\% and 43.0\% at 100 TeV and $8\times10^3$ TeV, respectively.
This improvement is more significant with increasing the primary energy. Fig. 3 shows that the energy resolutions at different zenith angles reconstructed
with the BNNs are obviously different from the LFM at 3000TeV. Compared to the LFM, the resolutions with the BNNs decrease by 48.5\% and 18.0\% at
zenith angles of $3^0$ and $43^0$, respectively. This improvement is more significant with decreasing the zenith angle. The energy resolutions are about 40\% and 18\% at
100 TeV and 1 PeV in the As-$\gamma$ and LHAASO experiments \cite{xuchen,lhaaso-ma}, respectively. The resolutions of 23.6\% and 13.4\% with the BNNs are obviously less than the ones
in the As-$\gamma$ and LHAASO experiments (see Fig. 2).
\end{flushleft}
\par
\begin{flushleft}
The difference between the results of the LFM and BNNs is just because the BNNs is a method of neural networks trained
by Bayesian statistics: first, the prior information explicitly,
which distinguishes the BNNs from the LFM, provides the necessary link
between the training sample and not yet predicted future
sample; second, the BNNs is a non-linear function like the neural
networks, and can extract more information from data than the LFM.
Therefore, the BNNs can be well applied to the energy reconstruction in
the EAS Experiments for ground-based TeV astrophysics, and the better energy resolution
can be obtained by the BNNs.
\par\end{flushleft}

\begin{flushleft}
Although the discussion in the present paper are only for the EAS
experiments, it is expected that the algorithm of the BNNs can also be
applied to the event reconstruction of other experiments and will
find wide application in the experiments of TeV astrophysics.
\par\end{flushleft}

\section{Acknowledgements }
This work was supported in part by the National Natural Science Foundation
of China (NSFC) under the contract No. 11235006, the Science Fund of
Fujian University of Technology under the contract No. GY-Z14061,
the Natural Science Foundation of Fujian Province in China under
the contract No. 2015J01577 and the Science and Technology Projects
of the Education Department of Fujian Province in China under the contract No. JB14071.


\newpage{}

\begin{figure}
\includegraphics[width=16cm,height=12cm]{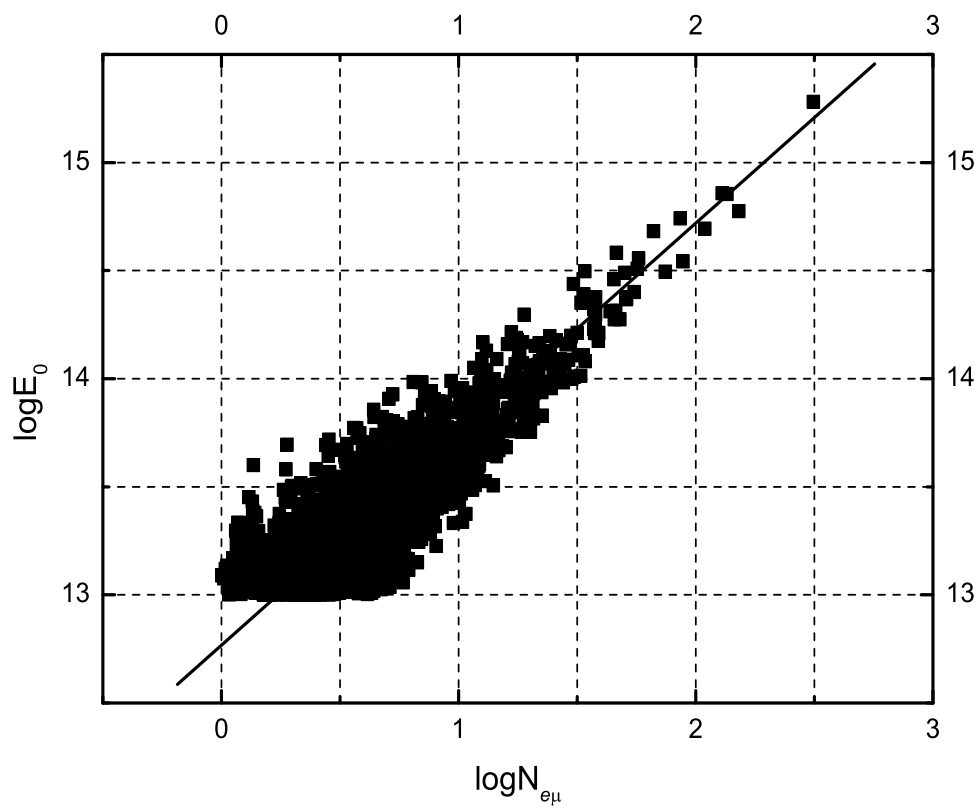}

\caption{The figure shows the approximative linear relationship between $\log{E_0}$ and $\log{N_{e\mu}}$.
The slope and intercept for the fitted line are 0.977 and 12.763, respectively.}
\end{figure}

\par

\begin{figure}
\includegraphics[width=16cm,height=12cm]{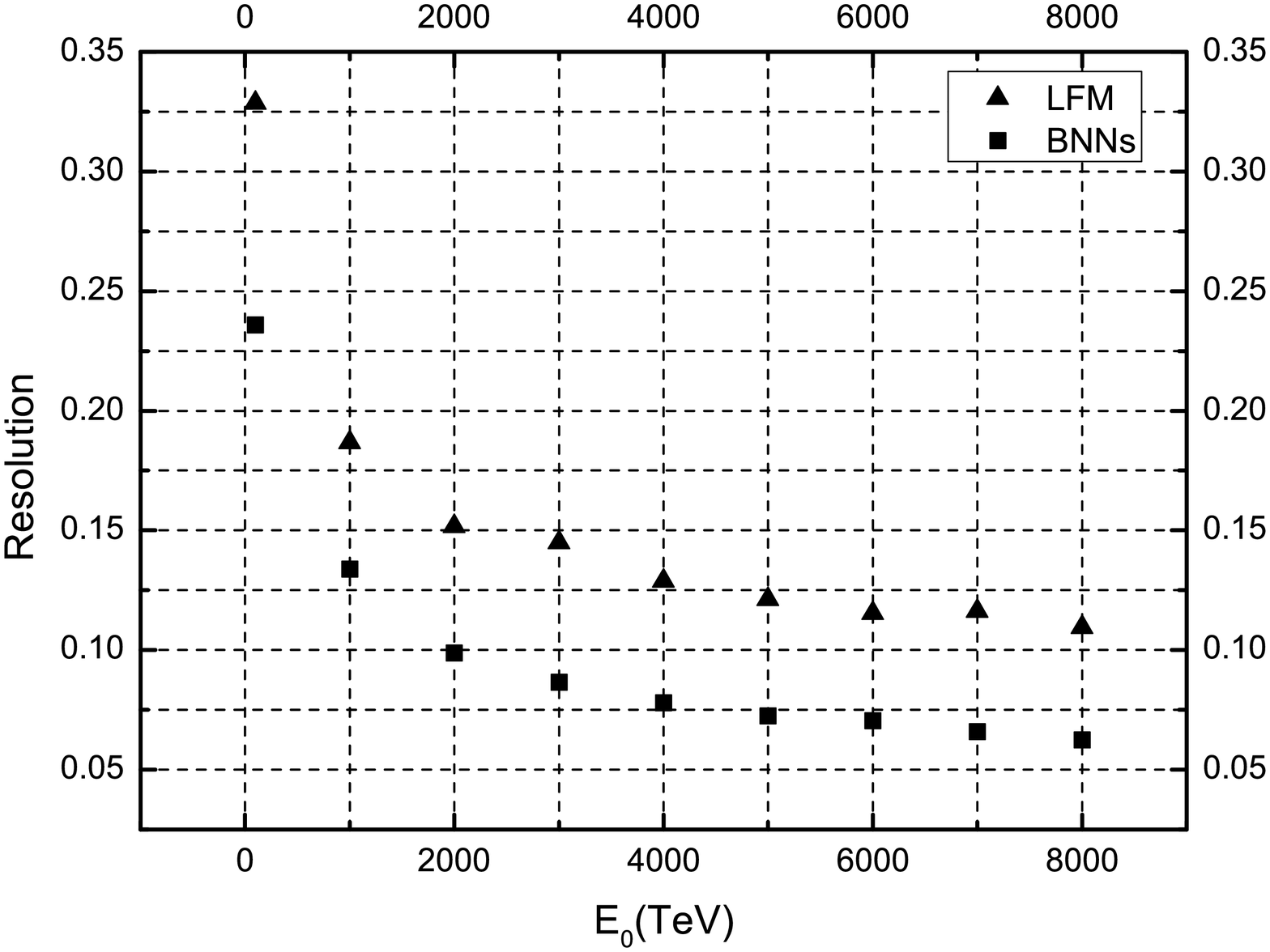}

\caption{The figure shows the energy resolutions reconstructed by the LFM and BNNs in the energy range from 100 TeV to $10^4$ TeV.
The black triangles and squares denote the resolutions from the LFM and BNNs, respectively.}
\end{figure}

\newpage{}
\begin{figure}
\includegraphics[width=16cm,height=12cm]{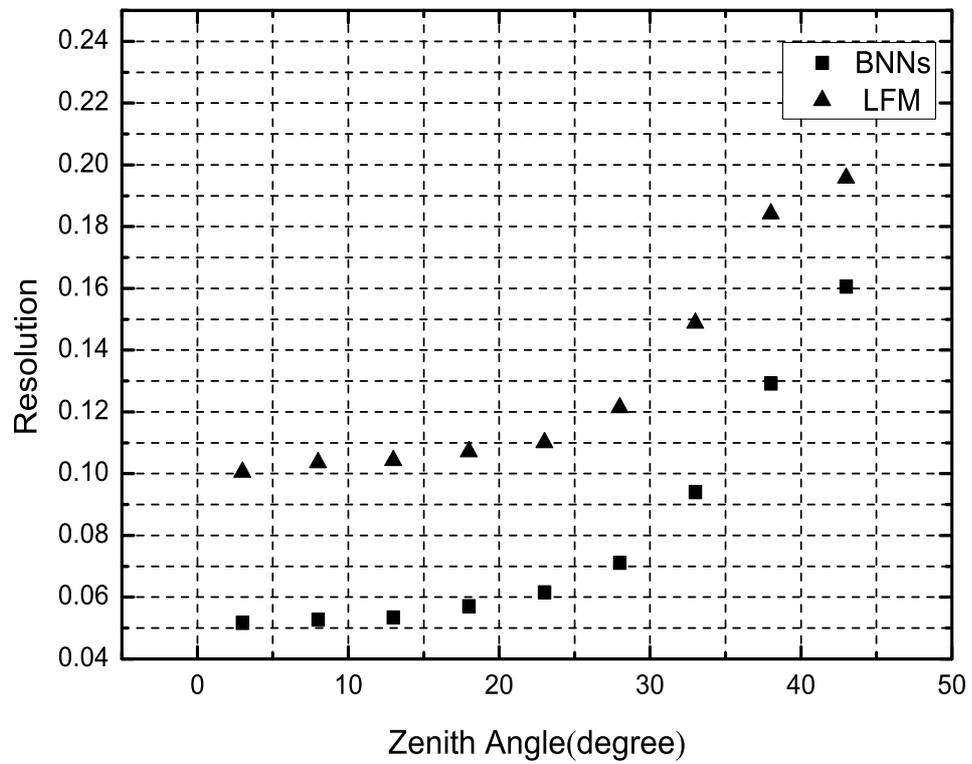}

\caption{The figure shows the energy resolutions reconstructed by LFM and BNNs in the zenith angle range from 0$^0$ to 45$^0$ at 3000 TeV.
The black triangles and squares denote the resolutions from LFM and BNNs, respectively.}
\end{figure}

\end{document}